\documentclass[aps,twocolumn,superscriptaddress]{revtex4-1}

\usepackage{amsmath}    
\usepackage{graphicx}   
\usepackage{verbatim}   
\usepackage{color}      
\usepackage{hyperref}   

\begin{document}

\title{Microscopic non-equilibrium energy transfer dynamics in a photoexcited metal/insulator heterostructure}

\author{N. Rothenbach}
\affiliation{Faculty of Physics and Center for Nanointegration
(CENIDE), University of Duisburg-Essen, Lotharstr.~1, 47057
Duisburg, Germany}

\author{M. E. Gruner}
\affiliation{Faculty of Physics and Center for Nanointegration
(CENIDE), University of Duisburg-Essen, Lotharstr.~1, 47057
Duisburg, Germany}

\author{K. Ollefs}
\affiliation{Faculty of Physics and Center for Nanointegration
(CENIDE), University of Duisburg-Essen, Lotharstr.~1, 47057
Duisburg, Germany}

\author{C. Schmitz-Antoniak}
\affiliation{Forschungszentrum J\"ulich, Wilhelm-Johnen-Str., 52428 J\"ulich, Germany}

\author{S. Salamon}
\affiliation{Faculty of Physics and Center for Nanointegration
(CENIDE), University of Duisburg-Essen, Lotharstr.~1, 47057
Duisburg, Germany}

\author{P. Zhou}
\affiliation{Faculty of Physics and Center for Nanointegration
(CENIDE), University of Duisburg-Essen, Lotharstr.~1, 47057
Duisburg, Germany}

\author{R. Li}
\affiliation{SLAC National Accelerator Laboratory, 2575 Sand Hill Rd., Menlo Park, California 94025, USA}

\author{M. Mo}
\affiliation{SLAC National Accelerator Laboratory, 2575 Sand Hill Rd., Menlo Park, California 94025, USA}

\author{S. Park}
\affiliation{SLAC National Accelerator Laboratory, 2575 Sand Hill Rd., Menlo Park, California 94025, USA}

\author{X. Shen}
\affiliation{SLAC National Accelerator Laboratory, 2575 Sand Hill Rd., Menlo Park, California 94025, USA}

\author{S. Weathersby}
\affiliation{SLAC National Accelerator Laboratory, 2575 Sand Hill Rd., Menlo Park, California 94025, USA}

\author{J. Yang}
\affiliation{SLAC National Accelerator Laboratory, 2575 Sand Hill Rd., Menlo Park, California 94025, USA}

\author{X. J. Wang}
\affiliation{SLAC National Accelerator Laboratory, 2575 Sand Hill Rd., Menlo Park, California 94025, USA}

\author{R. Pentcheva}
\affiliation{Faculty of Physics and Center for Nanointegration
(CENIDE), University of Duisburg-Essen, Lotharstr.~1, 47057
Duisburg, Germany}

\author{H. Wende}
\affiliation{Faculty of Physics and Center for Nanointegration
(CENIDE), University of Duisburg-Essen, Lotharstr.~1, 47057
Duisburg, Germany}

\author{U. Bovensiepen}\email[] {uwe.bovensiepen@uni-due.de}
\affiliation{Faculty of Physics and Center for
Nanointegration (CENIDE), University of Duisburg-Essen,
Lotharstr.~1, 47057 Duisburg, Germany}

\author{K. Sokolowski-Tinten}
\affiliation{Faculty of Physics and Center for Nanointegration
(CENIDE), University of Duisburg-Essen, Lotharstr.~1, 47057
Duisburg, Germany}

\author{A. Eschenlohr}\email[] {andrea.eschenlohr@uni-due.de}
\affiliation{Faculty of Physics and Center for
Nanointegration (CENIDE), University of Duisburg-Essen,
Lotharstr.~1, 47057 Duisburg, Germany}

\date{\today}

\begin{abstract} 

The element specificity of soft X-ray spectroscopy makes it an ideal tool for analyzing the microscopic origin of ultrafast dynamics induced by localized optical excitation in metal-insulator heterostructures. Using [Fe/MgO]$_n$ as a model system, we perform ultraviolet pump/soft X-ray probe experiments, which are sensitive to all constituents of these heterostructures, to probe both electronic and lattice excitations. Complementary ultrafast electron diffraction experiments independently analyze the lattice dynamics of the Fe constituent, and together with \textit{ab initio} calculations yield comprehensive insight into the microscopic processes leading to local relaxation within a single constituent or non-local relaxation between two constituents. Besides electronic excitations in Fe, which are monitored at the Fe L$_3$ absorption edge and relax within 1~ps by electron-phonon coupling, soft X-ray analysis identifies a change at the oxygen K absorption edge of the MgO layers which occurs within 0.5~ps. This ultrafast energy transfer across the Fe-MgO interface is mediated by high-frequency, interface vibrational modes, which are excited by hot electrons in Fe and couple to vibrations in MgO in a mode-selective, non-thermal manner. A second, slower timescale is identified at the oxygen K pre-edge and the Fe L$_3$ edge. The slower process represents energy transfer by acoustic phonons and contributes to thermalization of the entire heterostructure. We thus find that the interfacial energy transfer is associated with non-equilibrium behavior in the phonon system. Because our experiments lack signatures of charge transfer across the interface, we conclude that phonon-mediated processes dominate the competition of electronic and lattice excitations in these non-local, non-equilibrium dynamics. 

\end{abstract}

\maketitle

\section{Introduction}

Heterostructures provide access to well controlled material properties and allow for the design of new materials with the desired properties for electronic device and nano-scale transistor applications. They have led to a manifold of technological innovations in high-speed- and opto-electronics as well as in spintronics, developments which were awarded Nobel prizes to Kroemer and Alferov in 2000 and to Fert and Gr\"{u}nberg in 2007, respectively. Such artificial materials further promise functionality if combined with external stimuli, e.g. ultrafast laser pulses, leading to non-equilibrium dynamics. For example, quantized collective coherent phonons which stem from the phonon dispersion relation in the backfolded Brillouin zone of a semiconductor superlattice were identified \cite{bargheer_Sci_04}. All optical magnetization switching was found in artificial ferrimagnets \cite{mangin_14} and spin-current induced magnetization dynamics in heterostructures represent a rather recent development \cite{turgut_PRL_13, schellenkens_NC_14, razdolski_NC_17}.

In recent years considerable effort was made towards understanding the non-equilibrium properties of condensed matter with the goal of manipulating, and potentially controlling, material properties in response to an external, impulsive stimulus \cite{fleming_ratner_PhysToday_08}. This includes the analysis of elementary interaction processes on the microscopic femto- and picosecond timescales, at which individual steps of the quantum statistics of electronic and phononic excitations can be distinguished in a non-thermal regime. The multicomponent structure of heterostructures adds complexity in terms of interface vs. bulk effects as well as local vs. non-local dynamics (i.e. transfer of excitations and energy between the different constituents) to the already intricate problem of non-equilibrium dynamics. Therefore, an investigation of the non-equilibrium electron and lattice dynamics in such materials might be highly welcome; even more if such a study is specific to the constituents and their mutual interfaces.

For describing the interaction of the electron and lattice degrees of freedom during non-equilibrium dynamics, suitable assumptions facilitate treatment of the problem. The most prominent example is the two-temperature model (2TM) introduced by Anisimov et al., which describes the energy transfer from optically excited electrons in metals to the lattice empirically by a single electron-phonon coupling parameter \cite{anisimov_JETP_74, allen_PRL_87}. The excited electron and phonon distributions in this model are assumed to be at different electron and lattice temperatures $T_{\mathrm{e}}$ and $T_{\mathrm{l}}$, respectively. Taking the excitation density, the electronic and the lattice specific heat into account the equilibration of $T_{\mathrm{e}}$ and $T_{\mathrm{l}}$ is predicted to take up to few picoseconds. This model is widely applied because it is very handy and facilitates material specific predictions \cite{hohlfeld_ChemPhys_00}. However, it is questioned from theory whether it provides a realistic description for the actual non-equilibrium state \cite{baranov_PRB_14, kemper_AdP_17, weber_Arxiv_18}. The non-equilibrium dynamics of phonon distributions has been widely assumed as thermalized, likely due to the rather slow picosecond timescales involved. This might not be justified in general due to the rather weak, anharmonic coupling between phonons \cite{maldonado17, maldonado_submitted, chase_APL_16} and because energy transfer between electrons and phonons can drive relaxation dynamics in a non-thermal regime \cite{rameau_NatComm_16, Konstantinova_2018}. Experimental observations indicate that no simple general answer to this question exists.

The problem of energy transfer across interfaces is challenging by itself. The diffuse mismatch model is used widely to describe the thermal boundary or Kapitza resistance \cite{swartz_89} with extensions including optical phonons \cite{duda_10}, inelastic scattering \cite{huberman94, hopkins_09}, and interface roughness \cite{beechem_07}. By now it is clear that the strength of interface bonding as well as phonon dispersion and population are decisive \cite{monachon_16}. In case of metal-insulator interfaces an additional electronic contribution to the Kapitza conductance  may become important and lead to a considerable increase in the energy transfer due to scattering of electrons with interface vibrational modes \cite{huberman94,kst15}. However, it is so far not established how these energy transfer mechanisms evolve under strong non-equilibrium conditions, as they, like the 2TM, rely on describing the phonon system with a temperature. 

\begin{figure}
    \centering
    \includegraphics[width=0.99\columnwidth]{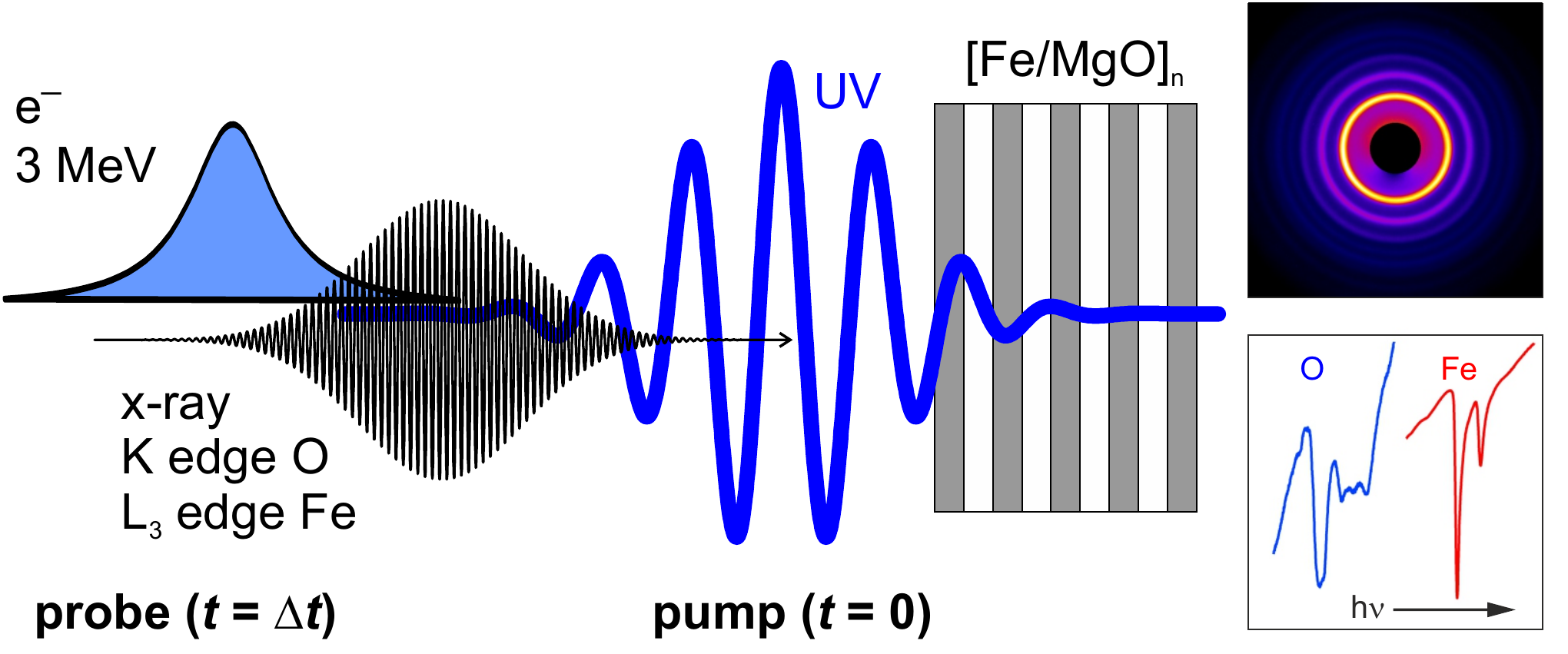}
    \caption{Schematic of UV pump, electron diffraction and soft x-ray absorption probe experiment. In both cases the signal transmitted through the [Fe/MgO]$_n$ heterostructure was detected. At right exemplary data are shown: Electron diffraction image in transmission with the diffraction intensity represented in a false color scheme (top) and x-ray transmission spectra (bottom) for the O K edge including a pre-edge feature originating from the Fe-MgO interface (blue) and the Fe L edges (red).}
    \label{fig:fig1}
\end{figure}

In this work, we investigate [Fe/MgO]$_n$ as a model system for a simple heterostructure and report on Fe site specific pumping combined with Fe and MgO specific probing with femtosecond time resolution. This approach provides direct insight into microscopic, dynamic redistribution of excitations and attendant energy transfer. X-ray absorption spectroscopy (XAS) at the O K- and the Fe L$_3$ edges combined with site specific information provided by the interface state in the O K pre-edge region allows one to analyze the non-equilibrium dynamics in the two constituents Fe and MgO as well as at their interface. We show by such solid state spectroscopy in a femtosecond (fs) time-resolved mode complemented by ultrafast electron diffraction, see Fig.~\ref{fig:fig1}, and \textit{ab initio} theory that the excess energy provided by ultraviolet excitation remains localized until coupling across the Fe-MgO interface. This coupling is found to be mediated by hot electrons in Fe which scatter with interface vibrations -- even for photon energies above the electron transfer gap. The identified dynamic process highlights the competition and predominance of local, inelastic electron-electron (e-e) scattering over charge transfer across interfaces in such heterostructures. Scattering of electrons in Fe with interface phonons is found to occur faster than thermalization in the initially excited Fe constituent and emphasizes the importance of non-thermal phonon distributions, which are concluded to be decisive for the fastest energy transfer dynamics among Fe and MgO.

\section{Sample preparation and characterization}

The investigated heterostructures [Fe/MgO]$_n$ with $n=1,4,5,8$ are sketched in Fig.~\ref{fig:fig1}. The samples were grown by molecular beam epitaxy at a sample temperature of $400$~K in a background pressure of $10^{-10}$~mbar, ensuring no contamination or oxidation of the individual deposited layers. The substrate consists of a 200~nm thick Si$_3$N$_4$ membrane, which carries a 100~nm thick Cu heat sink on its backside in order to dissipate the excess energy deposited by the pump beam. Although growth of Fe on MgO(001) can proceed epitaxially due to a rather good lattice match, the use of the Si$_3$N$_4$ membrane as a substrate leads to polycrystalline layer stacks. The individual Fe and MgO layers are, if not stated otherwise, 2~nm thick, which was monitored during growth by a quartz-crystal microbalance and subsequently determined by x-ray diffraction. In addition, the spatial extension of the interface was analyzed using interface sensitive Conversion Electron M\"{o}ssbauer Spectroscopy. From the obtained results we concluded that potential intermixing of the constituents is limited to one monolayer at the Fe-MgO interfaces, which are therefore considered to be atomically sharp. For more details on the sample and especially the interface quality, see the supplementary information.

\section{Results}

\begin{figure}
    \centering
    \includegraphics[width=0.99\columnwidth]{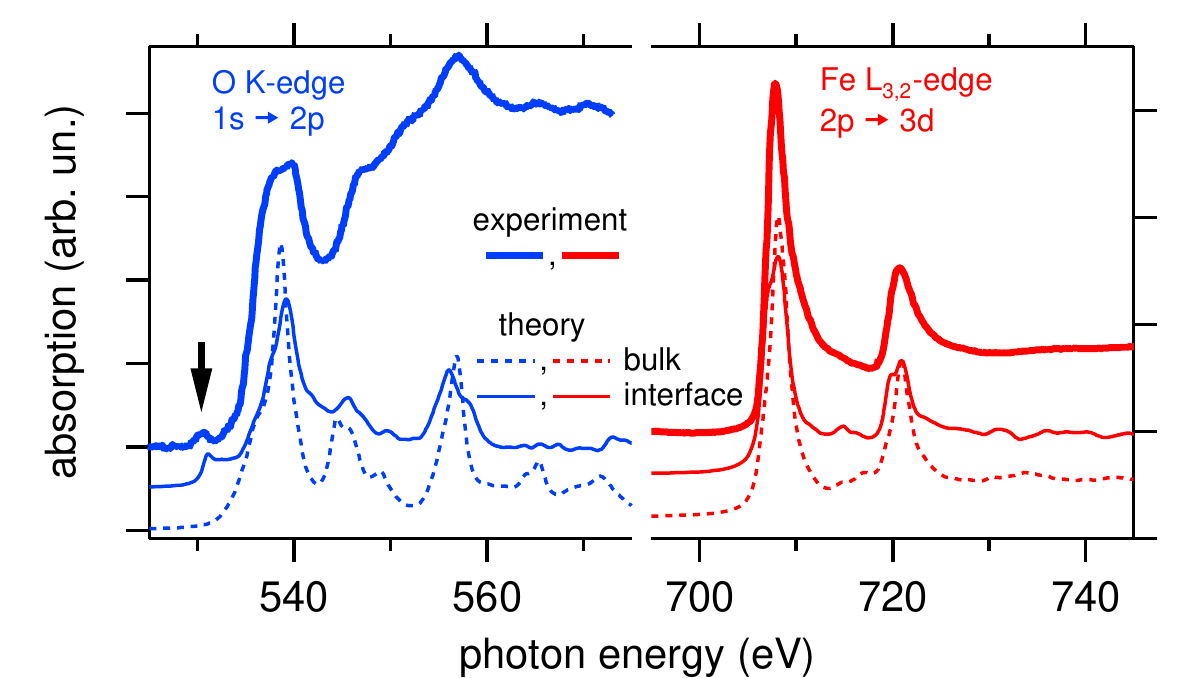}
    \caption{High resolution x-ray absorption spectra of [Fe/MgO]$_4$ with 2~nm individual layer thickness measured in transmission geometry (thick lines), normalized to the incident x-ray intensity. The Fe spectrum (red, at right) is composed of the L$_3$ and L$_2$ edges at 710 and 724~eV, respectively. The spectrum at the O K edge (blue, at left) contains several features in agreement with literature \cite{Luches_PRB_2004, Singh_Vacuum_2017, seabourne2010} and the vertical arrow points at the pre-edge feature assigned to an interface state, see text. Calculated atomic layer-resolved spectra of Fe$_8$/(MgO)$_8$ are shown by thin lines. Spectra at the interface are shown by a solid, those from the center of the respective layers by a dashed line. Spectra are shifted vertically for better visibility.}
    \label{fig:fig2}
\end{figure}


High resolution static soft x-ray absorption spectra of [Fe/MgO]$_4$, shown in Fig.~\ref{fig:fig2} for the O K edge and the Fe L$_{3,2}$ edges, were taken at the synchrotron light source BESSY II operated by Helmholtz Zentrum Berlin, beamline PM3 equipped with a plane grating monochromator and the ALICE endstation. Spectra taken at the Fe L$_{3,2}$ edges agree well with previous results for thin Fe films \cite{chen_PRL_1995}, those from the O K edge agree reasonably well with literature on thin MgO films on metal substrates \cite{Luches_PRB_2004}. There is a rather small pre-edge feature at 530~eV which was reported earlier at similar interfaces \cite{Luches_PRB_2004, Singh_Vacuum_2017, seabourne2010} and originates from an Fe-MgO interface state, as shown in the following by density functional theory (DFT) calculations. 

Here, the multilayer stack is represented by a periodically repeated heterostructure Fe$_8$/(MgO)$_8$(001) consisting of eight monolayers Fe and eight monolayers of MgO stacked along the (001) direction. Structural optimization and electronic density of states (DOS) was obtained with the VASP code \cite{cn:VASP1,cn:VASP2} using the generalized gradient approximation \cite{cn:Perdew96}. X-ray absorption spectra were calculated with the fully-relativistic SPRKKR Korringa-Kohn-Rostoker multiple scattering approach \cite{cn:SPR-KKR1,cn:SPR-KKR3}. (For comprehensive technical details see the supplemental material.) Fig.~\ref{fig:fig2} shows the calculated spectra for the O K edge (left) and the Fe L$_{3,2}$ edges (right), both in an atomic layer resolved manner as indicated. Comparing the experimental spectrum with the calculations, we see an overall agreement of the relative positions of the main features, including the small pre-edge hump at 530\,eV for the O K edge. The latter is clearly identified in the calculated spectrum of the interface layer while the spectrum calculated for the center of the MgO layer lacks this feature. Therefore, the pre-edge feature is assigned to an Fe-MgO interface state. This is in agreement with \cite{Luches_PRB_2004, seabourne2010}, which show that such a pre-edge feature occurs at the MgO-metal interface but not in bulk MgO. For Fe small interface induced changes obtained in the calculations were not resolved in experiment.

\begin{figure}
    \centering
    \includegraphics[width=0.99\columnwidth]{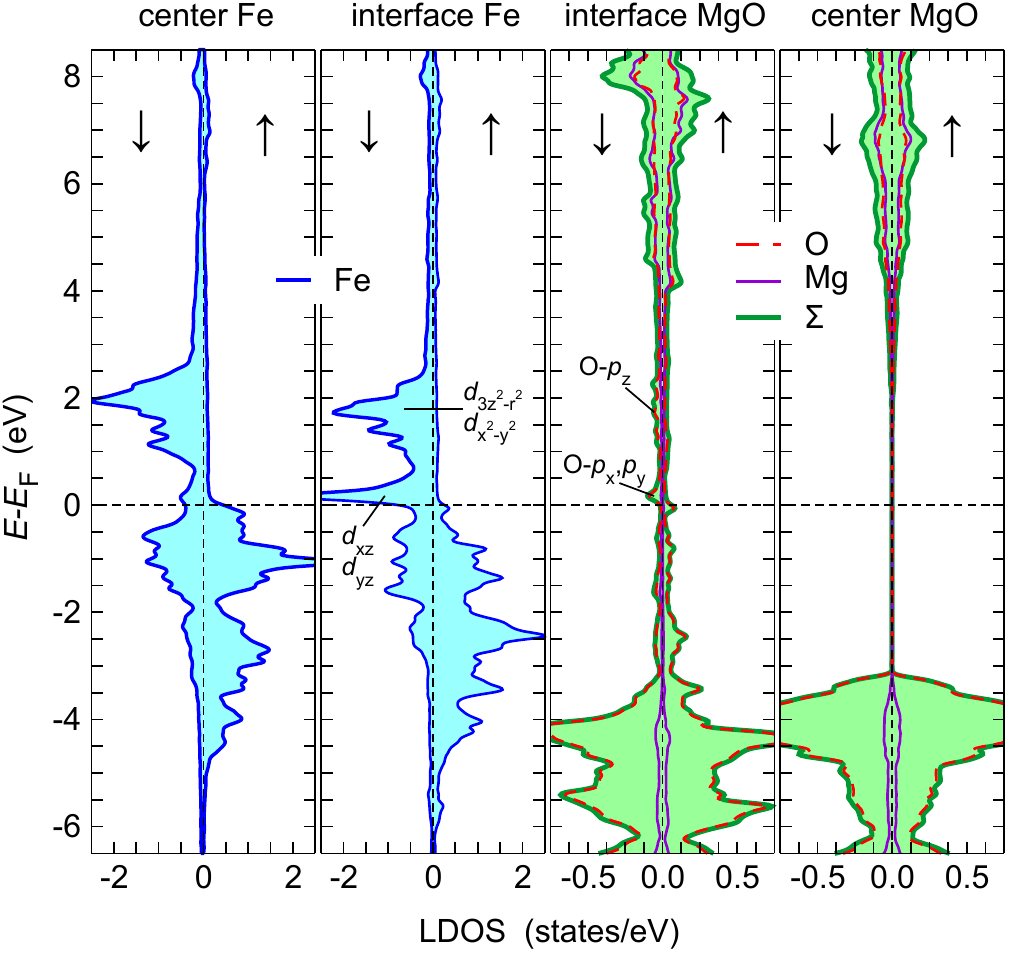}
    \caption{Spin-polarized element- and layer-resolved electronic density of states of Fe$_8$/(MgO)$_8$(001) obtained from DFT calculations. Only the interface layers (middle) and the central layers of the slabs (leftmost panel for Fe and rightmost for MgO) are shown. The contribution of Fe is represented by the blue lines, Mg by purple lines and O by the dashed red lines. The sum of Mg and O contributions is indicated by the green area. Upward and downward arrows denote the respective spin channels. While the central layers are similar to the respective bulk materials, pronounced features close to the Fermi energy appear in both interface layers arising from the strong hybridization between the Fe $d$-states and the $p$-states of the apical O. The DOS was calculated including the four intermediate layers. The result is shown in detail in the supporting material.}
    \label{fig:EDOS}
\end{figure}

The corresponding layer resolved DOS for majority and minority spin directions is shown in Fig.~\ref{fig:EDOS}, where we restrict the comparison to the central and interface layers of each part of the slab. While for Fe the Fermi-level crosses the upper edge of the majority $d$-band and passes through a wide valley in the minority channel, the central MgO layer can be considered insulating with its valence band maximum located 3.3\,eV below the Fermi energy $E_{\mathrm{F}}$. Strong modifications are present in the partial DOS in both the Fe and MgO interface layers. The hybridization of Fe and O orbitals leads to a considerable amount of interface states throughout the gap region. The most prominent feature is located in the minority channel approximately 0.2\,eV above the Fermi level. The large peak in the Fe interface layer has predominantly $d_{xz}$ and  $d_{yz}$ character and hybridizes with the $p_x$ and $p_y$ orbitals of the apical O. In turn, the features in Fe around 2\,eV consist of $e_g$-type orbitals, in particular $d_{3z^2-r^2}$, which leads to an increased $p_z$ character of the states at the interface O. The electronic structure of the heterostructure is thus expected to contain a relevant contribution of interfacial electronic states of Fe-MgO hybrid character.

\begin{figure}
    \centering
    \includegraphics[width=0.7\columnwidth]{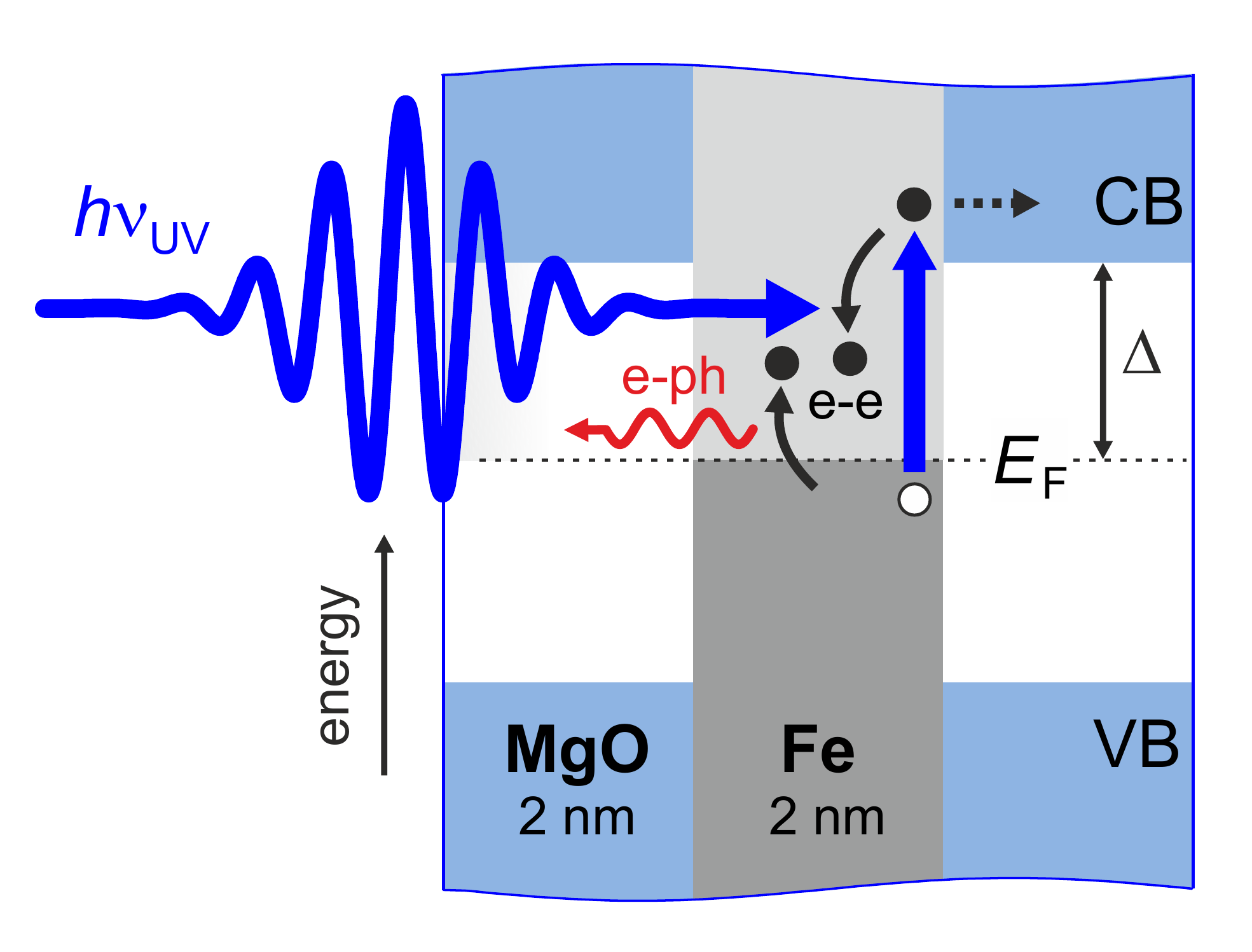}
    \caption{Excitation scheme at Fe/MgO interfaces. VB and CB indicate the MgO valence and conduction band, respectively. The excitation photon energy $h\nu_{\mathrm{UV}}=4.7$~eV, indicated by the vertical blue arrow, is generated by frequency tripling of the Ti:sapphire fundamental at 1.55~eV. $\Delta$ represents the charge transfer gap and is reported to be 3.8 eV \cite{petti_JPCS_11}. Relaxation processes due to electron-electron (e-e) scattering in Fe and electron-phonon (e-ph) scattering across the interface are indicated by solid arrows. Potential charge injection from hot electrons in Fe into the CB of MgO is depicted by a dashed arrow.}
    \label{fig:fig3}
\end{figure}

\subsection{Ultrafast soft x-ray spectroscopy}\label{sec:trXAS}

\begin{figure}
    \centering
    \includegraphics[width=0.99\columnwidth]{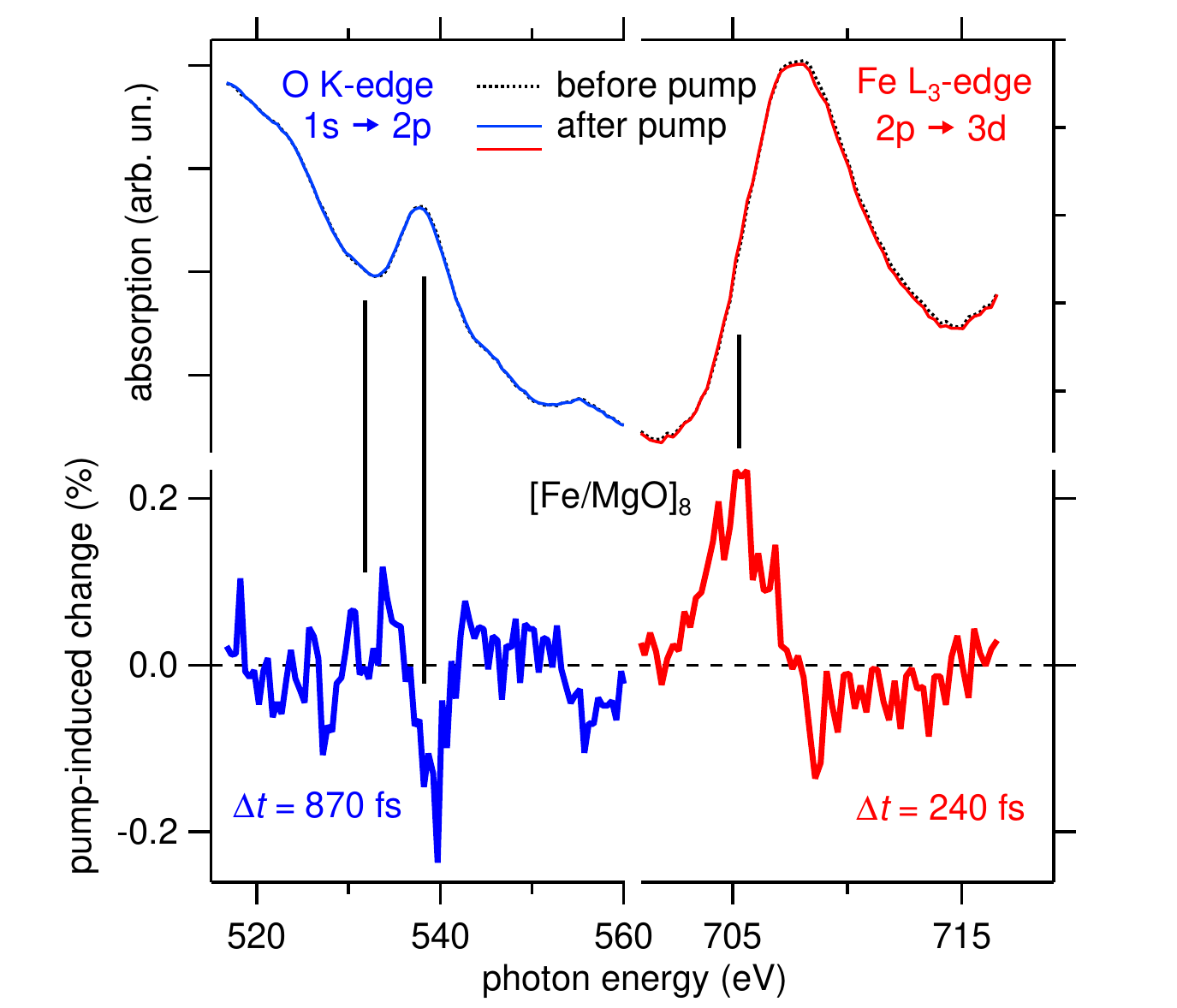}
    \caption{Top: Soft x-ray absorption as a function of photon energy for the oxygen K (blue) and the Fe L$_3$ edge (red) before and after pumping, measured at the FemtoSpex facility in transmission geometry \cite{XAS_normalization}. Bottom: UV pump-induced change in absorption as a function of soft x-ray photon energy at the indicated pump-probe delays of the maximum observed change. Vertical lines indicate photon energies at which time-dependent measurements were conducted.}
    \label{fig:fig4}
\end{figure}

By probing the Fe L$_3$ and O K edges we obtain element specific information. Since O occurs solely in MgO, we have chosen these absorption edges to obtain a constituent specific probe which we combine here with femtosecond time resolution \cite{bressler_Science_09, kachel_PRB_09, prendergast_PRL_11}. We performed pump-probe experiments at the FemtoSpex facility of BESSY II (beamline UE56/1-ZPM) \cite{holldack_JSR_14} and analyze the time-dependent response of the Fe and MgO constituents to ultraviolet pump excitations in the [Fe/MgO]$_n$ heterostructure. Pump photon energies above the electron transfer gap $\Delta$ at Fe-MgO interfaces provide an opportunity to investigate charge transfer excitations which transfer a hot electron optically excited upon absorption of an ultraviolet (UV) photon from Fe in the vicinity of $E_{\mathrm{F}}$ to the conduction band (CB) in MgO, see Fig.~\ref{fig:fig3}. The size of $\Delta = 3.8$~eV was reported in Ref.~\cite{petti_JPCS_11} to be close to the midgap position of the MgO band gap of 7.8~eV. Electron transfer across the interface is expected to compete with e-e scattering in the Fe layer. The latter is expected to be very efficient due to the large density of unoccupied $3d$ and $4s$ electronic states in Fe below the CB minimum of MgO.

Fig.~\ref{fig:fig4} shows in the top panels x-ray absorption spectra at the O K- and the Fe L$_3$ edge which were measured after UV pumping ($h\nu=4.7$~eV) with an incident fluence of $\approx 20$~mJ/cm$^2$ and $\approx150$~fs time resolution, derived from the pulse durations of the UV pump ($\approx50$~fs) and soft x-ray probe (100~fs \cite{holldack_JSR_14}). The temporal overlap of femtosecond UV pump and x-ray probe pulses was determined by an independent transmission experiment through a 20~nm thick Fe film, which exhibits a pump-induced change of the absorption at the L$_3$ edge similar to previous results for transition metal thin films \cite{Stamm_NatMater_2007, kachel_PRB_09}. Spectra at fixed pump-probe delays $t$ of the maximum pump-induced changes observed in [Fe/MgO]$_8$ at the O K edge, which occurs at 870~fs, and at the Fe L$_3$ edge at 240~fs are presented in the bottom panels. Note the different spectral shape of the pump-induced changes at the two absorption edges. The pump-induced change at the Fe L$_3$ edge consists of a first derivative-like spectral signature, which indicates a shift of the absorption edge, as reported for transition metals before \cite{kachel_PRB_09, prendergast_PRL_11}. The pump-induced change at the O K edge exhibits a different shape which is dominated by an intensity change at maximum absorption. Overall, the changes in absorption at the two edges are similar in strength up to 0.2\%, but occur at different time delays, which suggests that different physical processes are responsible for these changes.

Examples of pump-induced changes in XAS observed with 70~ps soft x-ray pulses are reported in the supplemental material in Fig.~S5. On such timescales electrons and phonons have equilibrated with each other and the heterostructure is close to its fully thermalized state \cite{hanisch_2012, maldonado17}. In this case the major part of the excess energy is hosted by phonons since the specific heat of the lattice is considerable larger then the one of the electrons. The observed changes showcase that time-resolved XAS is also sensitive to phonons, in agreement with temperature-dependent changes observed in literature \cite{kachel_PRB_09, nemausat_17}. XAS thus allows us to track the electronic and phononic energy transfer and relaxation in all constituents of the heterostructure.

\subsection{Ultrafast electron diffraction}\label{sec:UED}

To complement the time-resolved soft x-ray absorption spectroscopy experiments discussed above, we performed time-resolved ultrafast electron diffraction (UED) experiments to directly probe the lattice response of such heterostructures after fs laser excitation. These experiments were carried out using the MeV-UED facility at SLAC National Accelerator Laboratory \cite{weathersby15}, which provides ultrashort electron pulses at relativistic energies. The experiments reported here were carried out in normal incidence transmission geometry at a repetition rate of 120\,Hz with pulses of approx.\ ${\rm 2\times 10^5}$ electrons per pulse, a bunch duration of ${\rm \tau _{bunch} \approx}$ 200 fs FWHM, and a kinetic energy of $E_{\rm kin} = 3.7$~MeV. Due to the normal incidence diffraction geometry and MeV electron energy, resulting in a very short de Broglie wavelength, we probe essentially the in-plane lattice dynamics. Therefore, only the in-plane r.m.s. displacement ${\Delta\langle u^2\rangle}_p$ is measured. The quoted values for the total r.m.s. displacement ${\Delta\langle u^2\rangle}$ assume an isotropic response: ${\Delta\langle u^2\rangle}_p = \frac{2}{3} {\Delta\langle u^2\rangle}$.

To systematically address the role of interface effects as well as the importance of charge transfer processes, different sample configurations and pump photon energies have been investigated over an extended range of excitation fluences. Samples for UED comprised similar [Fe/MgO]$_n$-heterostructures as for the XAS experiments. However, in order to avoid additional background contributions to the overall scattering signal and due to the lower repetition rate of the experiment, a thinner, 20 nm thick Si$_3$N$_4$ membrane was used as a substrate without an additional metal layer as heat sink. The response of a [2~nm~Fe/5~nm~MgO]$_5$ multilayer heterostructure is compared to a [10~nm~Fe/25~nm~MgO]$_1$ bilayer using near normal incident pump pulses of approx.\ 50 fs duration at photon energies of 4.7 and 3.1~eV. Incident pump fluences of 6 to 15~mJ/cm$^2$ were employed.  While all interface mediated effects should be enhanced in the heterostructure compared to the bilayer, pumping in the UV with 4.7 eV photon energy allows electron transfer from Fe to MgO via a hot electron state. At 3.1~eV photon energy this process is suppressed, see Fig.~\ref{fig:fig3}. Compared to the soft x-ray experiments the relative amount of MgO in the samples has been increased to enhance the relatively weak scattering signal of MgO.

\begin{figure}
    \centering
    \includegraphics[width=0.99\columnwidth]{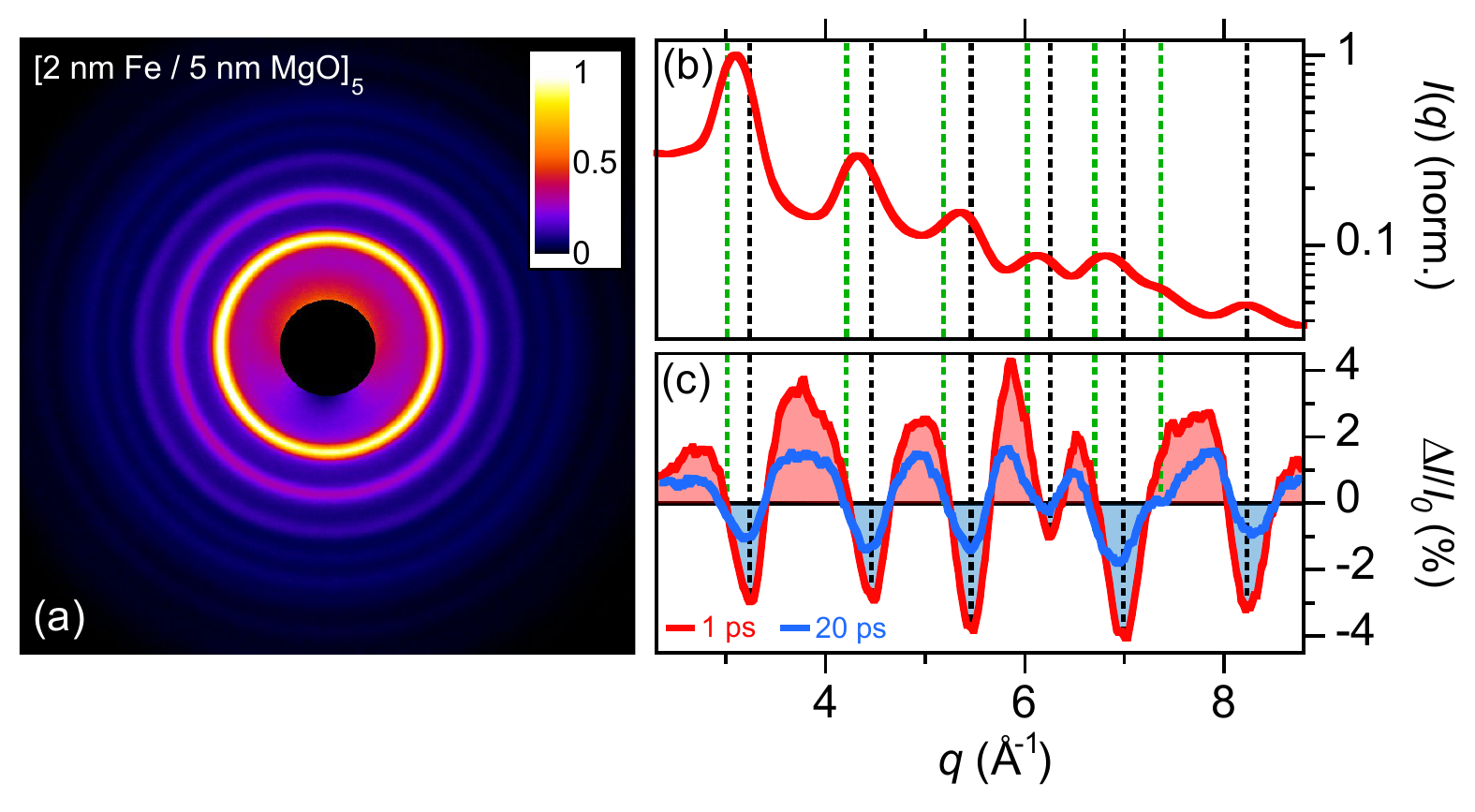}
    \caption{(a) Electron diffraction pattern of a [2~nm~Fe/5~nm~MgO]$_5$ heterostructure in false color representation. (b) Scattering intensity $I(q)$ as a function of momentum transfer $q$ of this heterostructure. (c) Difference scattering pattern $\Delta I\left(q,t\right) = I\left(q,t\right) - I_0\left(q\right)$, $I_0\left(q\right)$ represents the unpumped case, for delay times $t$=1~ps (red) and $t$=20~ps (blue) after 4.7~eV pump excitation at an incident fluence of $F$ = 9~mJ/cm$^2$. Black and green dashed vertical lines mark the positions of diffraction peaks of Fe and MgO, respectively.}
    \label{fig:fig6}
\end{figure}

As an example, Fig.~\ref{fig:fig6}a (left) depicts a diffraction image of [2 nm Fe/5 nm MgO]$_5$ without pumping. By azimuthal integration along lines of constant momentum transfer $q \approx 2\pi /\lambda\cdot \theta$, with de~Broglie wavelength $\lambda = 0.003$\, \AA \; and scattering angle $\theta$, the diffraction signal $I(q)$ is obtained, as displayed in Fig.\ \ref{fig:fig6}b. The diffraction intensity changes upon pumping as depicted in Fig.~\ref{fig:fig6}c, which shows the transient difference in the scattering intensity (pumped $-$ unpumped) for time delays $t=1$~ps (red) and 20~ps (blue), respectively, after excitation with 4.7~eV pulses at an incident fluence of $F$ = 9\,mJ/cm$\rm ^2$. A decrease of the Bragg peak intensities as well as an increase of the diffuse background inbetween the Bragg peaks is observed. Both features can be attributed to an incoherent excitation of the lattice and the increase of the r.m.s.\ atomic displacement after sample excitation also known as the transient Debye-Waller effect.

At the given momentum resolution of the experiment of 0.14\,\AA$^{\rm -1}$ the diffraction peaks of Fe and MgO overlap in most cases, as recognized from Fig.~\ref{fig:fig6}. Moreover, the transient changes are dominated by the response in Fe, in particular at earlier delay times, when the changes in MgO are relatively weak, c.f. the difference pattern at $t$ = 1 ps in Fig.\ \ref{fig:fig6}c. To quantitatively analyze the transient diffraction data the integrated signal of the individual Bragg-peaks has been determined by fitting them separately with a Gaussian function superimposed on a linear background for each time delay. Due to the overlap of diffraction peaks and the weak changes in MgO at early delays, we focus here on the results obtained for Fe.

The observed transient Debye-Waller effect describes the incoherent vibrational excitation of the lattice due to energy transfer from the optically excited electrons to phonons by the transient changes of r.m.s.\ atomic displacement ${\Delta\langle u^2\rangle}\left(t\right)$ according to

\begin{equation}
\label{eq:DWF}
-\ln\left(\frac{I_{hkl}\left(t\right)}{I^0_{hkl}}\right)=\frac{1}{3}\Delta\langle u^2\rangle (t) \cdot G^2_{hkl}
\end{equation}

Herein $I^0_{hkl}$ denotes the scattering signal of the unpumped sample measured at $t<0$, i.e. before the pump pulse excites the sample, $G_{hkl}$ the length of the reciprocal lattice vector corresponding to reflection $(hkl)$, and ${\Delta\langle u^2\rangle}$ the transient change of the r.m.s. displacement of ion cores. For a representative subset of the data we verified the dependence on diffraction order predicted by Eq.~\ref{eq:DWF} evidencing that within the experimental accuracy the lattice response in Fe is incoherent. However, for the bulk of the data we used the isolated (321)-peak at $q=8.19$~\AA$^{-1}$ to determine ${\Delta\langle u^2\rangle}(t)$, since this peak is not influenced by diffraction from MgO.

\subsection{Time-resolved results of soft x-ray absorption and electron diffraction}\label{sec:comparison}

\begin{figure}
    \centering
    \includegraphics[width=0.99\columnwidth]{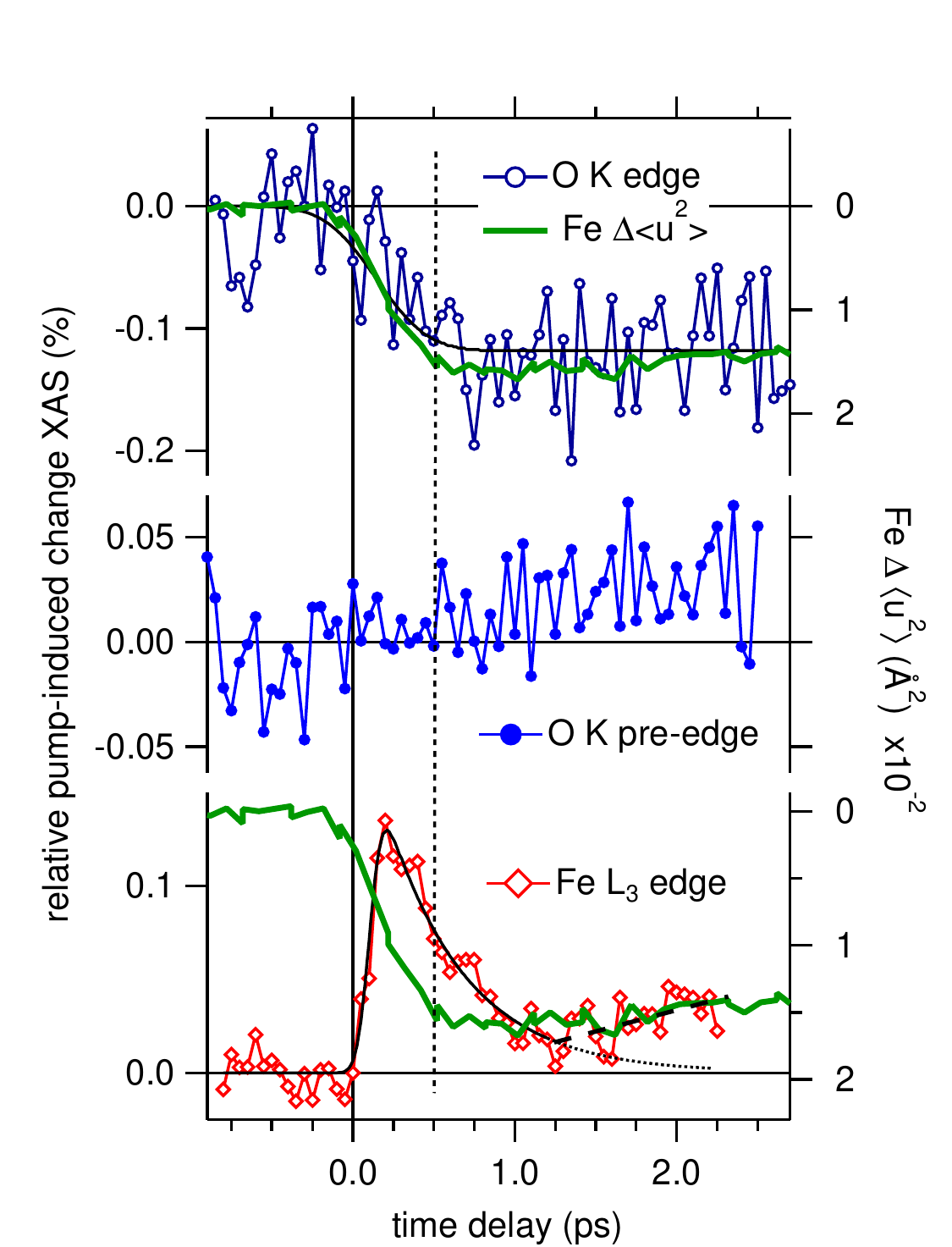}
    \caption{Pump-induced changes with 4.7~eV pump photon energy observed for [Fe/MgO]$_8$ at three selected soft x-ray photon energies (left axis) in combination with r.m.s. atomic displacement (right axis) as a function of pump-probe delay. Top: Maximum absorption of the O K edge, and r.m.s. displacement of the Fe ion core positions determined from changes in the electron diffraction intensity dominated by Fe ion cores. Center: Pre-edge region of O K edge at the Fe-MgO interface state. Bottom: Fe L$_3$ edge and r.m.s. atomic displacement (identical to the top panel). Black lines are fits to the data as specified in \cite{fit_XAS}. The dashed and dotted lines guide the eye, see text.}
    \label{fig:fig5}
\end{figure}

Employing two complementary ultrafast methods on equivalent heterostructures now allows us to conclude on the ultrafast electron and lattice dynamics. We show the observed transient changes jointly in Fig.~\ref{fig:fig5}. For analysis of the pump-induced dynamics the soft x-ray photon energy was kept fixed at three selected values indicated by vertical lines in Fig.~\ref{fig:fig4}. These energies are at (i) the maximum absorption of the O K edge, where also the maximum pump-induced change is observed, (ii) the corresponding pre-edge region, and (iii) at the Fe L$_3$ edge at the energy of maximum change, i.e. 2~eV below the maximum absorption. The pump-induced changes were measured as a function of pump-probe delay and the results are shown in Fig.~\ref{fig:fig5} by blue and red symbols. The transients at the different energies exhibit a very different behavior. The vertical dashed line at 0.5 ps highlights this fact. At this delay the change at the Fe L$_3$ edge has already gone through the maximum and has started to recede. The change at the O K edge has just reached its maximum and the change in the O K pre-edge region is still building up. Time zero of the independent XAS and UED experiments is determined as the time delay at which the pump-induced changes begin, with a precision of $\pm50$~fs given the finite probe pulse durations.

In more detail, we observe at the Fe L$_3$ edge (Fig.~\ref{fig:fig5}, bottom panel) a fast increase in absorption which relaxes almost back to zero change until 1.2~ps. For a fixed x-ray photon energy at the rising flank of the Fe L$_3$ edge, the transient shift of the spectrum to lower photon energies (cf. Fig.~\ref{fig:fig4}) will indeed manifest as an absorption increase. At later delays a second positive change appears, however with a slower time constant. At the O K edge (top panel) a decrease in absorption builds up within 0.5~ps and stays constant up to 3~ps. At the O pre-edge (center panel) we observe an increase in absorption which continues to increase within the investigated time delay. The time-dependent intensity at the O K edge was fitted with a step function, while at the Fe L$_3$ edge, an exponential relaxation with a time $\tau_2$ was taken into account in addition. To account for the build up of the change with time constant $\tau_1$, both functions were convoluted with a Gaussian before fitting the data. A time shift $t_0$ with respect to the experimentally determined time zero, which is defined as the onset of the transient change, was taken into account in addition. This fit function was chosen for the minimum number of free parameters needed to describe the data, and without assuming a particular physical model, see \cite{fit_XAS}. The fit shown in Fig.~\ref{fig:fig5}, bottom panel, was done for $t<1.2$~ps and describes the data well with $\tau_1=0.14\pm0.05$~ps and $\tau_2=0.49\pm0.06$~ps. At later delay times the fit, which is extrapolated to reach the zero level at 2.5 ps (dotted line), does no longer describe the data. At these delay times the second absorption increase is highlighted by the dashed line. We note that such an absorption increase is observed for the heterostructure, but not for a single 20~nm polycrystalline Fe film as shown in Fig.~S4 of the supplemental material. Also data for single films of Ni reported previously \cite{Stamm_NatMater_2007,kachel_PRB_09} do not exhibit such a second intensity increase. We therefore assign this second, slower intensity increase to the presence of interfaces. We further note that the dynamics at the O K edge occur exclusively in the heterostructure, i.e. are mediated by a transfer of excitations from Fe. In a single MgO reference layer, no transient changes are observed at the O K edge after pumping, see Fig.~S3 in the supplemental material.

\begin{figure}
    \centering
    \includegraphics[width=0.99\columnwidth]{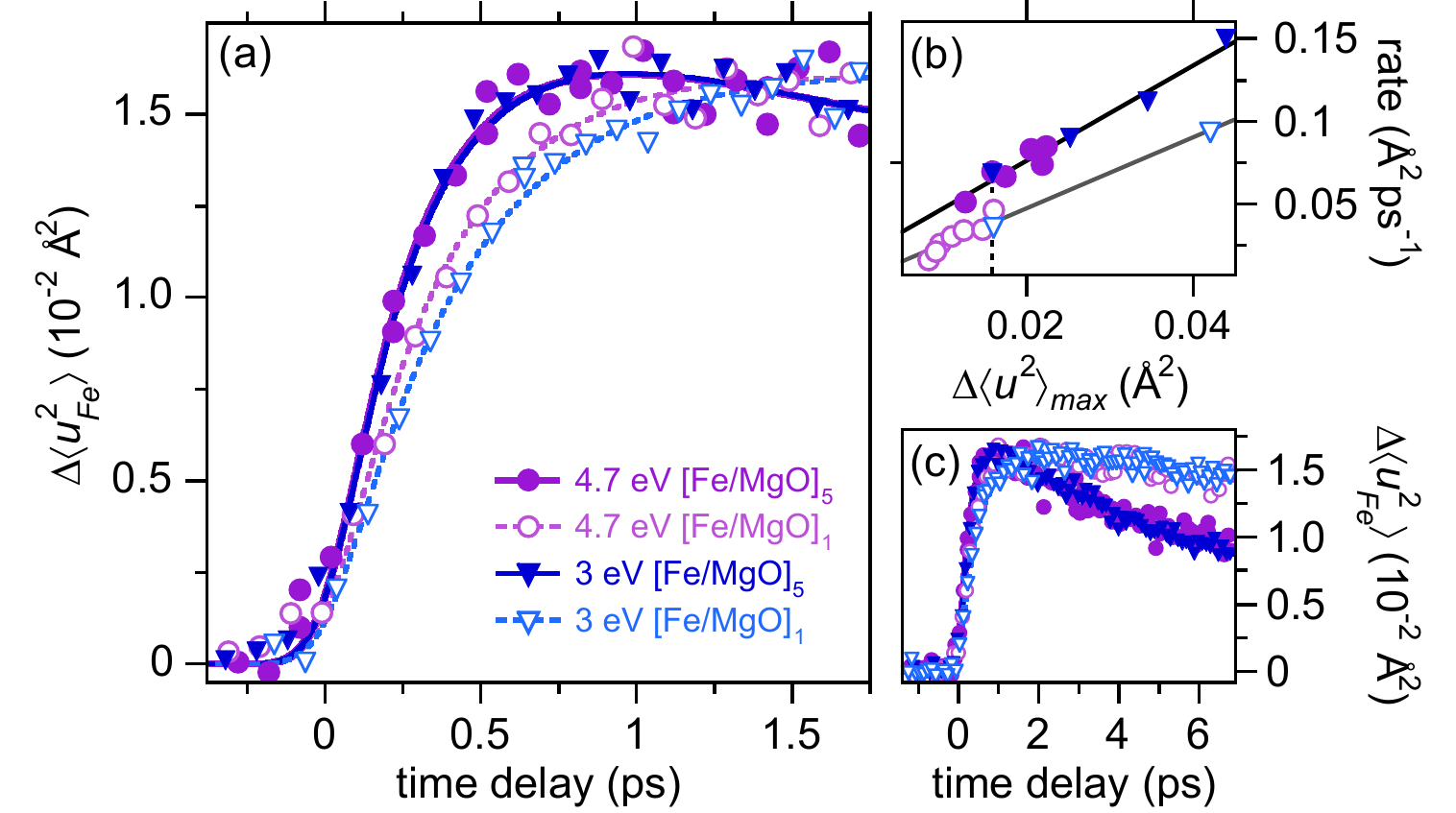}
    \caption{(a): Change of the r.m.s.\ displacement ${\Delta\langle u^2 \rangle}(t)$ as a function of pump-probe time delay for the different sample configurations ([2~nm~Fe/5~nm~MgO]$_5$ - filled symbols; [10~nm~Fe/25~nm~MgO]$_1$ - open symbols) and excitation conditions (pump photon energy 4.7~eV - circles, 3.1~eV - triangles); the solid and dashed curves represent fits to the measured data as explained in the text. (b) Maximum rate of the increase of ${\Delta\langle u^2\rangle}(t)$ obtained from the fits as shown in panel (a) for a range of pump fluences. The different values of maximum ${\Delta\langle u^2\rangle}$ were obtained from pump fluences ranging from 6 to 15~mJ/cm$^2$. Photon energies are indicated by symbols, see panel (a). Solid curves: Linear guide to the eye. (c) ${\Delta\langle u^2\rangle}(t)$ over a larger delay time range.}
    \label{fig:fig7}
\end{figure}

Regarding the UED results, Fig.~\ref{fig:fig7}a compares the obtained ${\Delta\langle u^2\rangle}(t)$, which exhibit the same maximum ${\Delta\langle u^2\rangle}_{max} \approx$ 0.016 \AA$^{\rm 2}$  for the different multi- and bilayer samples, filled and open symbols, respectively, and excitation conditions ($h\nu=$4.7~eV - circles; 3.1 eV - triangles). The data show that the excitation of the lattice occurs within a few hundred femtoseconds indicative of strong electron-phonon coupling in Fe. We emphasize that the response exhibits clear differences for multi- and bilayers, but is independent of the pump photon energy.

To allow a quantitative analysis these data have been fitted by

\begin{eqnarray}
	\label{eq:fit_u_squared}
  \Delta\!\left<u^2_\mathrm{Fe}\right>\!(t) & \!\propto\! &\int_{-\infty}^{\infty}\!\!\!d\tau\,\,\,e^{-\frac{(t-\tau)^2}{2\sigma^2}} \quad \times \\\nonumber
  & \!\times\! & \left[A_\mathrm{r}\left(1-e^{-\frac{\tau}{\tau_\mathrm{r}}}\right)-
      \Theta(\tau-\tau_\mathrm{r})A_\mathrm{d}\left(1-e^{-\frac{\tau-\tau_\mathrm{r}}{\tau_\mathrm{d}}}\right)\right]
\end{eqnarray}

i.e. by an exponential rise (amplitude $A_{\mathrm{r}}$, time constant $\tau_{\mathrm{r}}$) followed by an exponential decay (amplitude $A_{\mathrm{d}}$, time constant $\tau_{\mathrm{d}}$) convoluted with a Gaussian function of 200~fs FWHM to account for the finite electron pulse duration \cite{epulse}. The resulting fits are shown in Fig.\ \ref{fig:fig7}a by solid and dashed curves. Fig.~\ref{fig:fig7}b presents the maximum rate of the increase of ${\Delta\langle u^2\rangle}\left(t\right)$ as a function of the maximum increase of the r.m.s.~displacement ${\Delta\langle u^2\rangle}_{max}$ as derived from the amplitudes $A_{\mathrm{r}}$ and time-constants $\tau_{\mathrm{r}}$ of the initial exponential rise of the fits shown in Fig.~\ref{fig:fig7}a as well as from fits of similar time dependencies measured over an extended range of excitation fluences, corresponding to ${\Delta\langle u^2\rangle}_{max}$ from 0.008 \AA$^2$ to 0.042 \AA$^2$, see Fig.~\ref{fig:fig7}b. Consistently and independent of pump photon energy all time dependencies measured for the heterostructure multilayer sample (filled symbols) exhibit a faster rise of ${\Delta\langle u^2\rangle}$ compared to the bilayer sample (open symbols). The subsequent relaxation exhibits also pronounced differences between the multilayer and the bilayer as demonstrated by Fig.\ \ref{fig:fig7}c, which shows ${\Delta\langle u^2\rangle}\left(t\right)$ from the same measurements as in Fig.\ \ref{fig:fig7}a over an extended pump-probe delay range. While in the multilayer the decay occurs within 5 ps, the corresponding time constant is almost an order of magnitude larger in the bilayer. This difference is explained by interface effects which contribute to an accelerated lattice response in Fe. As discussed in the following section, we conclude that combined hot electron- and phonon-mediated processes at the interface are essential.




\section{Discussion}

We start with observations related to the Fe constituent, which is primarily excited by the pump pulse, and compare the time-dependent results of x-ray absorption and electron diffraction experiments. The time-dependent x-ray absorption at the rising flank of the Fe L$_3$ edge is characterized by an absorption increase, which reaches its maximum at 200~fs, see Fig.~\ref{fig:fig5}, and results from a red shift of the Fe L$_3$ absorption edge, cf. Fig.~\ref{fig:fig4}. This observation is in qualitative agreement with literature results for Ni films \cite{Stamm_NatMater_2007, carva_EPL_2009, kachel_PRB_09}. Since UV absorption in Fe occurs within the pump pulse, the fact that the maximum x-ray absorption increase is reached well after the pump pulse is explained by inelastic electron-electron (e-e) and hole-hole scattering leading to relaxation of the primarily excited, non-equilibrium electron distribution and redistribution of electrons and holes towards $E_{\rm F}$ \cite{anisimov_JETP_74,hohlfeld_ChemPhys_00}. Eventually the electrons thermalize at an electron temperature $T_{\rm e}$ higher than the initial static temperature $T_0$. As the timescale of the initial rise corresponds to typical electron thermalization times in transition metals \cite{rhie_03}, we explain the Fe L$_3$ edge red shift by such electron redistribution in the vicinity of $E_{\rm F}$, leading to increased absorption below and reduced absorption above $E_{\rm F}$. We note that such redistribution is characteristic for a metal where the chemical potential is positioned within an electron band. Relaxation of the absorption increase is assigned to dissipation of the excess energy in the electronic system mediated by coupling to bosons \cite{rameau_NatComm_16}. The resulting lattice dynamics are analyzed by the reported ultrafast electron diffraction, which probes electron-phonon (e-ph) coupling by an increased mean square displacement $\langle u^2\rangle$ of predominantly Fe ion cores, see Sec. \ref{sec:UED}, Fig.~\ref{fig:fig7}.

We further compare the thicker Fe films with the [Fe/MgO]$_8$ layer stack regarding the time-dependent x-ray absorption and r.m.s. displacement. We find that ${\Delta\langle u^2\rangle}$ reaches its maximum for the heterostructure at 1~ps, while it peaks at 2~ps for the [10~nm~Fe/25~nm~MgO]$_1$ film, see Fig.~\ref{fig:fig7}. This matches the minima in the time-dependent x-ray absorption reached for the heterostructure at 1~ps, see Fig.~\ref{fig:fig5}. For a comparison with XAS on a 20~nm thick Fe reference sample see Fig.~S4 of the supplemental material. The time delay of the minima in the time-dependent Fe L$_3$ absorption, which indicates dissipation of the electronic excess energy by coupling to phonons, matches well with the maxima in ${\Delta\langle u^2\rangle}$, which probes the phonon excitation. Thus, our combined absorption and diffraction study verifies the energy transfer from the optically excited electrons to the lattice. Crucially, both experimental methods identify faster dynamics in the heterostructure compared to the bulk film and thus probe the interface mediated coupling at identical time scales.

While the electron diffraction data exhibit a comparatively simple dynamics with two time scales for increase and decrease in ${\Delta\langle u^2\rangle}$, the time-dependent behavior of the absorption at the Fe L$_3$ edge exhibits, in addition to the initial increase and decrease, a third time scale indicating a second, delayed process with increasing absorption for $t>1.2$~ps, see Fig.~\ref{fig:fig5}. At such time delays ${\Delta\langle u^2\rangle}$ is decreasing, which represents energy dissipation out of the Fe lattice, and we conclude that at these later times the heterostructure is thermalizing as a whole.

\begin{figure}
    \centering
    \includegraphics[width=0.99\columnwidth]{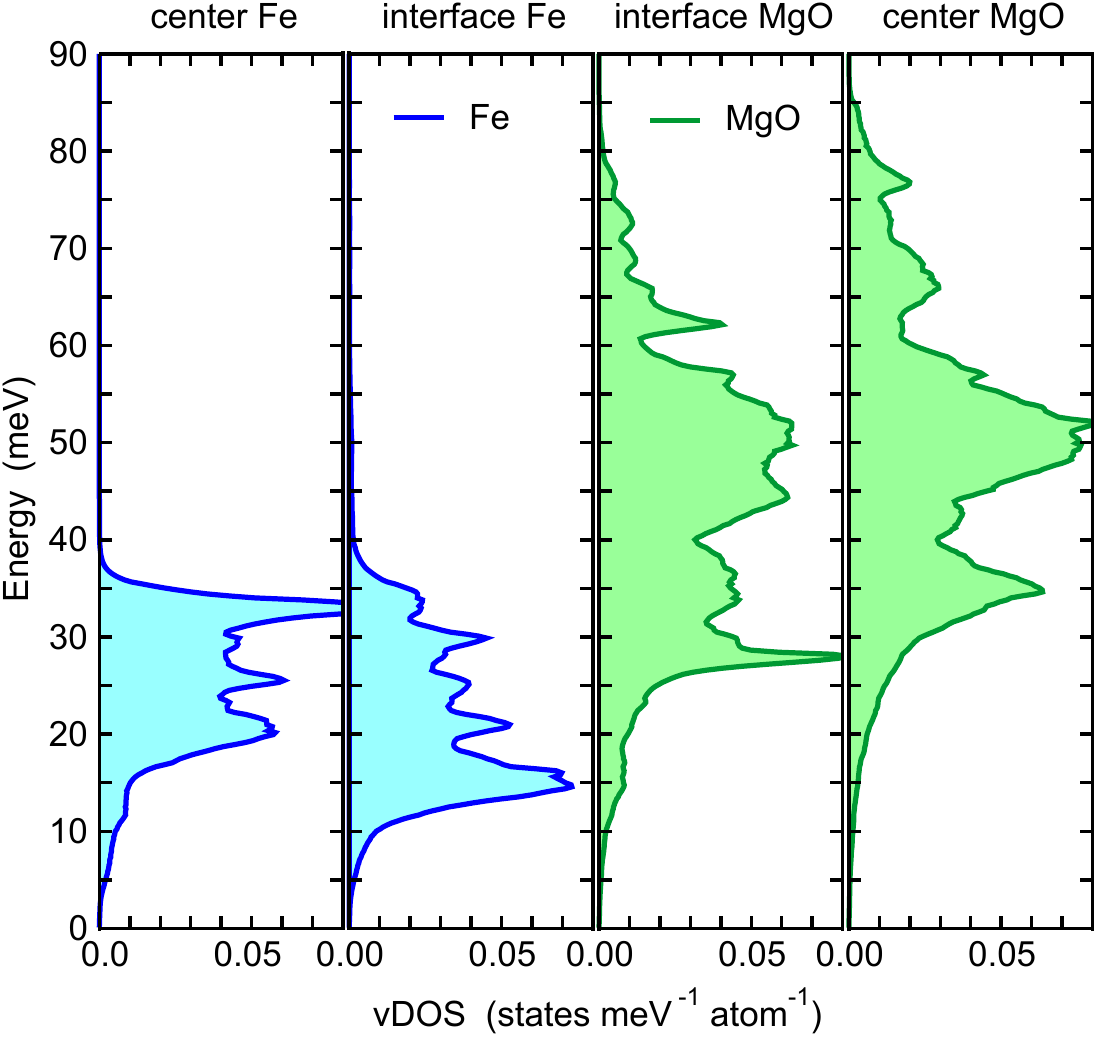}
    \caption{Layer-resolved vibrational density of states of Fe$_8$/(MgO)$_8$(001) obtained from DFT calculations. Colors and symbols as in Fig.\ \protect\ref{fig:EDOS}. Again, we show only the central layers of the slabs (leftmost panel for Fe and rightmost for MgO) and the interface layers (middle). The entire VDOS including the four intermediate layers is provided in the supplementary material.}
    \label{fig:VDOS}
\end{figure}

We will come back to this aspect further below and turn our attention now to the time-dependent absorption at the O K edge, which probes the oxide constituent. We observe an absorption decrease which saturates within 500~fs, see Fig.~\ref{fig:fig5}, top. Remarkably, this transient behavior matches within experimental uncertainties well with the transient increase in ${\Delta\langle u^2\rangle}$ representing Fe ion cores, Fig.~\ref{fig:fig5}, top, green line. This observation implies that the Fe lattice is excited in parallel with the MgO constituent. The O K edge absorption change represents transfer of excitations from Fe to MgO. The fact that the maximum change in the oxide is reached at later delays compared to the Fe L$_3$ edge indicates that this effect is not primarily related to the transfer of hot electrons to MgO which should occur on a significantly shorter time scale, as investigated in \cite{gruner_19}. This conclusion agrees with our observation in time-resolved UED that the lattice dynamics in the heterostructure is independent of the pump photon energy (see Fig.~\ref{fig:fig7}) and can be understood as follows. The lifetime of an electron in Fe at $E-E_{\mathrm{F}}>3.8$~eV which could potentially transfer elastically from Fe into MgO, as illustrated in Fig.~\ref{fig:fig3}, is only several fs \cite{bauer_PSS_15} and its ultrafast relaxation is mediated by inelastic e-e scattering in Fe. Formation of a hot electron temperature occurs within 200~fs \cite{rhie_03}. Therefore, local e-e scattering in Fe competes with hot electron transfer to MgO. Even if an electron is transferred to MgO it would gain an energy $\Delta$ by a transfer back to Fe and relaxation to $E_{\mathrm{F}}$, see Fig.~\ref{fig:fig3}. All our experimental results indicate consistently that e-e scattering in Fe dominates this competition.

From the observed timescale of the time-resolved XAS at the O K edge we thus derive that transfer of the excitations from the metal to the insulator involves phonons.  Fig.~\ref{fig:fig5} shows on the one hand that lattice excitations in Fe and MgO build up simultaneously. On the other hand, energy flow from Fe to MgO must take place since the initial hot electron population remains localized in Fe. This energy transfer requires coupling of either hot, thermalized electrons in Fe to hybrid phonon modes at the Fe-MgO interface or coupling of these hot electrons to bulk phonons in Fe which couple to phonons in MgO. The latter mechanism is widely considered in the diffuse mismatch model \cite{monachon_16}, in which a temperature gradient drives the energy transfer across the interface described by a thermal boundary conductance. Since the timescale related with this mechanism is in the range of about 10~ps or longer \cite{hanisch_2012} and we observe the increase of the oxygen signal before e-ph equilibration in Fe occurs, we exclude a thermal mechanism as an explanation for the ultrafast excitation of MgO. As discussed by Huberman and Overhauser \cite{huberman94}, on the metallic side electrons can couple to hybrid phonons which are localized at the metal-insulator interface and decay into both constituents. In a thermal picture, this process determines an electronic Kapitza conductance. It was found earlier that such coupling is essential for other heterostructures \cite{kst15,kst17} in which the faster relaxation in thinner samples with two instead of only one interface has been attributed to a coupling of hot electrons in the metal to interface vibrational modes \cite{huberman94,sergeev98}. Accordingly, energy transfer to hybrid phonon modes localized at the interface represents an additional channel for hot electrons in Fe to relax, as long as the electron and lattice systems in Fe have not yet equilibrated. The layer resolved vibrational density of states (vDOS) of [Fe/MgO]${_n}$ calculated by DFT (see the supplementary material for the theoretical details) and reported in Fig.~\ref{fig:VDOS} provides insight in the phonon modes contributing to this process. At energies above the acoustic phonons, which have a rather small vDOS,  there is sizeable vDOS for the center and interface layer, see Fig.~\ref{fig:VDOS}. Our calculations indicate that at the interface at 27~meV a peak in the vDOS occurs in MgO. For the center MgO layer this peak is found at 35~meV. Also Fe exhibits sizable vDOS at these energies in the bulk and at the interface with an interface mode at 15~meV which couples to MgO as indicated by a hump in the interface vDOS of MgO at the same energy. This combination of high vDOS in both constituents of the heterostructure at the interface corroborates the scenario that the energy transfer across the interface within 1~ps is predominantly mediated by coupling of hot electrons to interface phonons at energies at 15~meV and above. These interface phonons decay into bulk. Due to the small film thickness of few nm of the heterostructure's constituents their vibrational amplitude is not localized at the interface but reaches well into the individual layer.

\begin{figure}
    \centering
    \includegraphics[width=0.99\columnwidth]{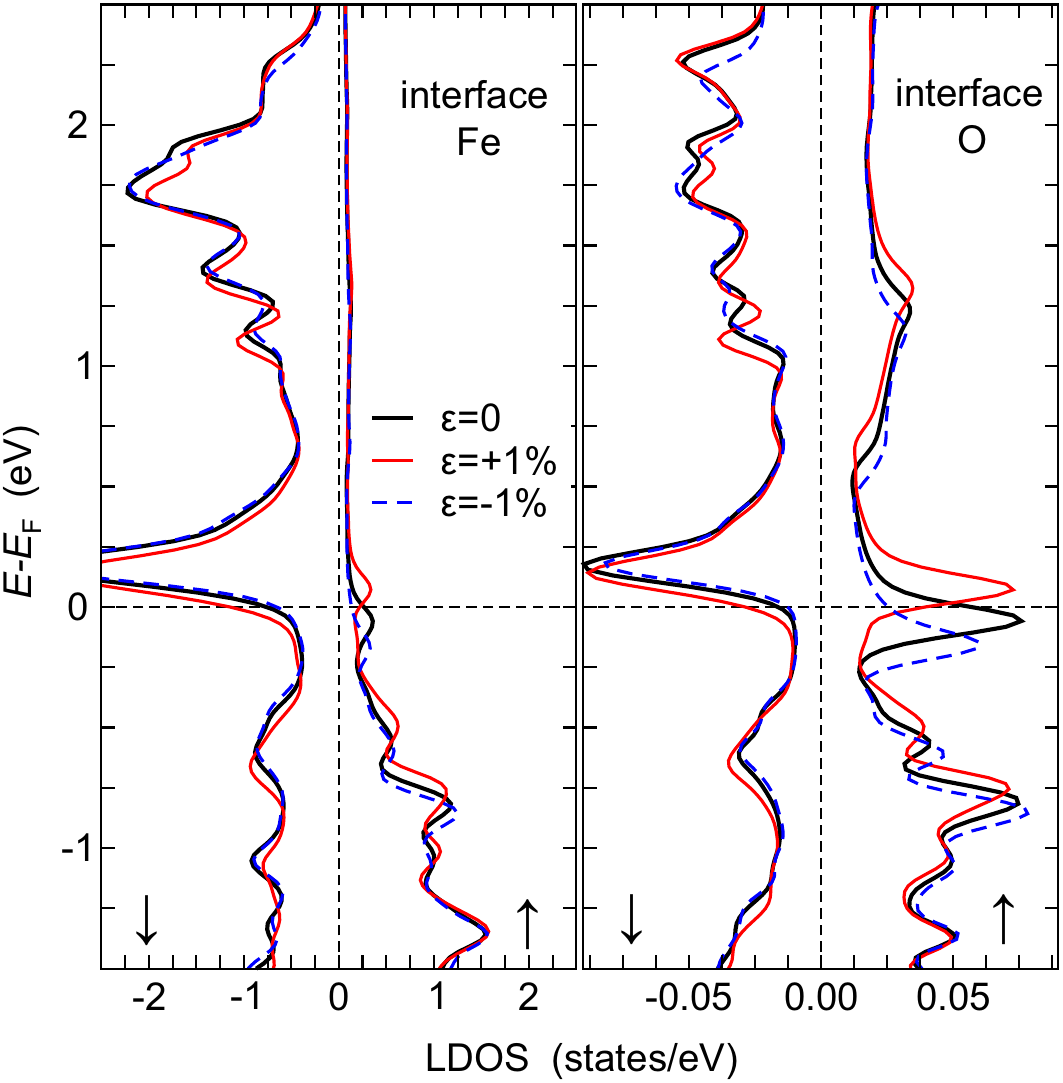}
    \caption{Layer-resolved electronic density of states under 1\% compression (red line) and expansion (blue, dashed line) of Fe$_8$/(MgO)$_8$(001) at the Fe-MgO interface obtained from DFT calculations. 1\% compression and expansion maintain the hybrid character of the interface states. The strongest response to lattice compression/expansion is found for majority electronic states ($\uparrow$) of the interface at O sites, first panel from the right, with shifts of up to 0.1~eV. Similar shifts are observed for the Fe majority interface states, while the minority states in both Fe and O shift by about 0.03~eV per 1\% compression/expansion.}
    \label{fig:strain}
\end{figure}

We now shift our attention to longer time delays. For 1~ps~$<t<5$~ps the receding ${\Delta\langle u^2\rangle}(t)$ observed in Fig.~\ref{fig:fig7}c indicates decay of Fe vibrations in the heterostructure [Fe/MgO]$_5$, while in the single Fe film  [Fe/MgO]$_1$ the maximum occurs later and the relaxation is weaker. This different behavior occurs because in the heterostructure the MgO constituents act as sinks for the vibrational excess energy in Fe, which is supported by a simple argument from the conversion of ${\Delta\langle u^2\rangle}$ into a lattice temperature \cite{thermalization}. In XAS we observe increasing signals at these later delay times at the Fe L$_3$ edge and at the O K pre-edge, see Fig.~\ref{fig:fig5}. As shown in the bottom panel the Fe signal grows at $t>1$~ps concomitantly with the O K pre-edge intensity in the center panel. The O K pre-edge feature originates from hybridized states of Fe and the apical O at the interface, which are located close to the Fermi level, see Fig. ~\ref{fig:EDOS}. These are in particular, the Fe $3d_{3z^2-r^2}$ and O $2p_z$ orbitals in the majority spin channel and the Fe $3d_{xz}$ and $3d_{yz}$ and O $2p_x$ and $2p_y$ orbitals in the minority spin channel. This energetic position of the hybridized states makes them very sensitive to local lattice distortions, which we have modelled by uniaxial lattice compression and expansion. The results are reported in Fig.~\ref{fig:strain} and show that the electronic interface state responds to lattice compression/expansion while the center layers barely show a change. Our experimental results obtained at the O K pre-edge therefore suggest that on longer timescales $t>1$~ps a local lattice distortion at the interface builds up. This effect is expected as a consequence of the different static thermal expansion coefficients of MgO and Fe \cite{Lehr_1956, Dewaele_2000}. According to the calculations reported in Fig.~\ref{fig:strain}, such an expansion shifts the position of the interface state and affects the resonant absorption at the O K pre-edge. We anticipate a similar origin for the observed change at the Fe L$_3$ edge at $t>1$~ps. However, here features in the near-edge fine structure related to interfacial hybridization between Fe and O were not spectrally resolved in the experiment, compare Fig.~\ref{fig:fig2}.

We conclude that two different mechanisms of phonon-mediated energy transfer act in such heterosystems. The faster energy transfer process at $t<1$~ps is mediated by high frequency interface phonons excited by hot electrons in Fe, in analogy to an electronic Kapitza conductance, but during phonon non-equilibrium. The slower transfer process at $1<t<5$~ps is closer to a thermal limit and involves lower energy acoustic phonons. MgO is characterized by ionic bonding with a partially covalent character. In particular, the interface states are based on Fe $d$- and O $p$-orbitals oriented along defined directions and high frequency modes. Coupling of Fe electrons or phonons with such high frequency modes likely proceeds on considerably faster time scales compared to acoustic phonons in metals \cite{hohlfeld_ChemPhys_00,ligges_APL_09} or semiconductors \cite{shah_99}, as has been observed in previous work on graphite \cite{kampfrathPRL05} and cuprates \cite{perfettiPRL_06,rameau_NatComm_16,Konstantinova_2018}. In particular for heterostructures, mode specific coupling of non-thermal phonons may offer new opportunities to selectively transfer energy among constituents before phonon thermalization occurs in a time range of 10-100~ps \cite{maldonado17}.

\section{Conclusions and Outlook}

The performed complementary pump-probe experiments analyze electronic and phononic contributions to dynamical processes in metal-insulator heterostructures [Fe/MgO]$_n$ on ultrafast timescales after localized optical excitation in Fe. The pump-induced change in the soft x-ray absorption at the Fe L$_3$ edge peaks at 200~fs pump-probe delay, which indicates electronic excitations localized in Fe. Further indications of electronic excitation were not observed in the experiments, which highlights the competition of hot electron transfer from Fe to MgO across the interface and electron-electron scattering in Fe. We conclude that the latter process dominates and is essential in understanding the subsequent dynamics in the lattice of the heterostructure. The soft x-ray absorption change at the O K edge, which probes the MgO constituent, saturates at 500~fs simultaneously with the increased mean square displacement of Fe ion cores probed in ultrafast electron diffraction. This indicates coupling of hot electrons in Fe with an interfacial phonon mode. Layer-resolved calculations of the vibrational density of states find a pronounced overlap of high frequency phonons above 15 meV phonon energy between Fe and MgO at the interface. This leads us to conclude on (i) a strong and energy-selective coupling across the interface and (ii) an initially highly non-thermal, hot phonon population in MgO. Both pump-probe experiments identify at time delays $t>1$~ps a second, slower timescale, which is attributed to thermalization of the heterostructure as a whole and energy transfer across the interface involving lower energy phonons.

Access to energy selective phonon excitation across interfaces opens opportunities to manipulate transient material properties by specific vibrations, i.e. beyond lattice heating. As demonstrated here, heterostructures allow to spatially separate excitation and material response and offer means to optimize the respective constituents towards desired properties.

\begin{acknowledgments}

We acknowledge fruitful discussions with J. J. R. Rehr and financial support by the Deutsche Forschungsgemeinschaft (DFG, German Research Foundation) through the Colloborative Research Center (CRC) 1242 (project number 278162697, TP A05, C01, C02), FOR 1509 (project WE2623/13) as well as CRC/TRR 247 (project B2). Calculations were carried out on the MagnitUDE supercomputer system (DFG grants INST 20876/209-1 FUGG, INST 20876/243-1 FUGG) of the Center for Computational Sciences and Simulation at the University of Duisburg-Essen. Helmholtz Zentrum Berlin is gratefully acknowledged for the allocation of synchrotron radiation beamtime and for financial support for travel to BESSY II. The UED work was performed at SLAC MeV-UED, which is supported in part by the DOE BES SUF Division Accelerator \& Detector R\&D program, the LCLS Facility, and SLAC under contract Nos. DE-AC02-05-CH11231 and DE-AC02-76SF00515. We thank N. Pontius, R. Mitzner, K. Holldack, C. Sch\"{u\ss}ler-Langeheine, E. W\"ust, A. Smekhova, N. Svechkina and R. Abrudan for experimental support. We further like to thank U. von H\"orsten for his expert technical assistance with sample preparation.

\end{acknowledgments}

\bibliography{rothenbach_PRX,me}

\begin{thebibliography}{61}
\bibitem[{XAS()}]{XAS_normalization}
\bibinfo{note}{Note that for technical reasons, namely the orders of magnitude
  lower x-ray flux compared to regular synchrotron operation, normalization of
  these x-ray absorption spectra to the incident x-ray intensity was not
  possible. This leads to a sloping background. Moreover, the spectra are
  instrumentally broadened due to a reduced energy resolution compared to the
  PM3 beamline.}
\bibitem[{fit()}]{fit_XAS}
\bibinfo{note}{The time-dependent XAS signal at the O K edge is fitted by
  $\Delta I_{\mathrm{O K}}(t) \propto
  \Theta(t,t_0)\otimes\exp{-\frac{(t-t_0)^2}{\tau^2}}$ and the one at the Fe
  L$_3$ edge by $\Delta I_{\mathrm{Fe L_3}}(t) \propto
  \left[\Theta(t,t_0)\cdot\exp{-\frac{t-t_0}{\tau_2}}\right]\otimes
  \exp{-\frac{(t-t_0)^2}{\tau_1^2}}$, $\Theta$ is the Heaviside function.}
\bibitem[{epu()}]{epulse}
\bibinfo{note}{The electron pulse duration has not been explicitely measured in
  this experiment. However, the fitting parameters and thus the derived rate of
  the increase of ${\Delta\langle u^2\rangle}$ showed only minor changes when
  varying the electron pulse duration over a range from 150 fs to 300 fs. Best
  fitting (judged from the onset of ${\Delta\langle u^2\rangle}$ at early delay
  times) was obtained for a pulse duration of 200 - 250 fs.}
\bibitem[{the()}]{thermalization}
\bibinfo{note}{While we are aware that even for simple metals the phonon system
  exhibits initially a non-thermal population \cite{chase_APL_16,maldonado17},
  conversion of the measured ${\Delta\langle u^2\rangle}$ into changes of the
  lattice temperature ${\Delta T}$ is instructive to characterize the flow of
  incoherent vibrational energy ("heat") in the system. Using published
  temperature dependent Debye-Waller factors for Fe (e.g.\ \cite{kharoo77} and
  references therein) the maximum increase of ${\Delta\langle u^2\rangle}_{max}
  \approx$ 0.016 \AA$^2$ (data shown in Fig.\ \ref{fig:fig7}a, c) corresponds
  to a peak temperature of $T_{\mathrm{Fe,max}} = (610 \pm 15)$K. For the
  heterostructure and both, 4.7~eV and 3.1~eV pumping, the r.m.s.\ displacement
  of Fe decreases with a time-constant of about 5~ps to a quasi-stationary
  value of ${\Delta\langle u^2\rangle} \approx$ 0.0056 \AA$^2$ corresponding to
  an asymptotic temperature of the Fe lattice of $T_{Fe,\infty} = (410 \pm 10)$
  K. At these late time delays it is also possible to derive a value for the
  Debye-Waller factor in MgO (using the (422)-reflection at $q =
  7.29$~\AA$^{-1}$). Using again published temperature dependent Debye-Waller
  factors \cite{baldwin64, singh82} the derived ${\Delta\langle u^2\rangle}
  \approx 0.0041$~\AA$^2$ yields a temperature of $T_{\mathrm{MgO},\infty} =
  (430 \pm 10)$~K, indicating that Fe and MgO have equilibrated at a common
  temperature. Note that this temperature is higher for smaller MgO
  thicknesses, and therefore the temperature drop from the maximum lattice
  temperature to the value several ps later is smaller. Keeping the relatively
  large noise in the O K edge tr-XAS experiment in mind, this might explain why
  we do not observe a relaxation of the pump-induced change at the O K edge
  within the analyzed delay range.}
\end{thebibliography}

\end{document}